# A Survey on Recognition-Based Graphical User Authentication Algorithms

Farnaz Towhidi
*Centre for Advanced Software Engineering,*
University Technology Malaysia
Kuala Lumpur, Malaysia
farnaz.towhidi@gmail.com

Maslin Masrom
*College of Science and Technology*,
University Technology Malaysia
Kuala Lumpur, Malaysia
maslin@ic.utm.my

*Abstract*—Nowadays, user authentication is one of the important topics in information security. Strong text-based password schemes could provide with certain degree of security. However, the fact that strong passwords are difficult to memorize often leads their owners to write them down on papers or even save them in a computer file. Graphical authentication has been proposed as a possible alternative solution to text-based authentication, motivated particularly by the fact that humans can remember images better than text. In recent years, many networks, computer systems and Internet-based environments try used graphical authentication technique for their user's authentication. All of graphical passwords have two different aspects which are usability and security. Unfortunately none of these algorithms were being able to cover both of these aspects at the same time. In this paper, we described eight recognition-based authentication algorithms in terms of their drawbacks and attacks. In the next section, the usability standards from ISO and the related attributes for graphical user authentication usability are discussed. The related attack patterns for graphical user authentication security part are also discussed. Finally, a comparison table of all recognition-based algorithms is presented based on ISO and attack patterns standards.

*Keyword-Recognition-Based Graphical User Authentication, Graphical Password, Usability, Security, ISO usability, Attack Pattern.*

I. INTRODUCTION

In recent years, computer and network security has been formulated as a technical problem. A key area in security research is authentication which is the determination of whether a user should be allowed access to a given system or resource. In this context, the password is a common and widely authentication method still used up to now.

A password is a form of secret authentication data that is used to control access to a resource. It is kept secret from those not allowed access, and those wishing to gain access are tested on whether or not they know the password and are granted or denied access accordingly.

The use of passwords goes back to ancient times when soldiers guarding a location by exchange a password and then only allow a person who knew the password. In modern times, passwords are used to control access to protect computer operating systems, mobile phones, auto teller machine (ATM) machines, and others. A typical computer user may require passwords for many purposes such log in to computer accounts, retrieving e-mail from servers, accessing to files, databases, networks, web sites, and even reading the morning newspaper online.

Some drawbacks of normal password appear like stolen the password, forgetting the password, and weak password. Therefore, a big necessity to have a strong authentication method is needed to secure all our applications as possible. Traditionally, conventional passwords have been used for authentication but they are known to have security and usability problems. Today, other method such as graphical authentication is one of the possible alternative solutions.

Graphical password have been proposed as a possible alternative to text-based, motivated particularly by the fact that humans can remember pictures better than texts. Psychological studies have shown that people can remember pictures better than text. Pictures are generally easier to be remembered or recognized than text, especially photos, which are even easier to be remembered than random pictures [17].

In graphical password, the problem arises because passwords are expected to have two fundamentals requirements, namely

a) Password should be easy to remember.
b) Password should be secured.

Graphical passwords were originally described by Blonder [5]. In his description, an image would appear on the screen, and the user would click on a few chosen regions of it. If the correct regions were clicked in, the user would be authenticated. Memorize ability of password and efficiency of their inputs is two key human factors criteria. Memorize ability have two aspects, that is,

a) How the user chooses and encodes the password?
b) What task the user does when retrieving the password?

In a graphical password system, a user needs to choose memorable image. The process of choosing memorable images depends on the nature of the process of image and the specific sequence of click locations. In order to support memorize ability, images should have meaningful content because meaning for arbitrary things is poor.

II. GRAPHICAL PASSWORDS METHODS

In this section, some graphical password systems based on recognition and recall-based are discussed. Graphical-based password techniques have been proposed as a solution to the conventional password techniques because graphic pictures are more easily remembered than texts which most of researchers have nominated them as "Picture superiority effect" [18].

A literature on most of articles regarding graphical password techniques from 1994 till January 2009 shows that the techniques can be categorized into three groups as below.

*A. Recognition-Based Technique*

In this category, users will choose pictures, icons or symbols from a collection of images. In authentication process, the users need to recognize their registration choice among a set of candidates. The research shows that 90% of users can remember their passwords after one or two month [15].

*B. Pure Recall-Based Technique*

In this category, users need to reproduce their passwords without being given any reminder, hints or gesture. Although this category is easy and convenient, but it seems that users hardly can remember their passwords similar to DAS (1999) and Qualitative DAS (2007).

*C. Cued Recall-Based Technique*

In this category, the technique proposed a framework of reminder, hints and gesture that help the users to reproduce their passwords or help users to make a reproduction more accurate similar to Blonder Algorithm (1996) and Passpoint (2005).

III. RECOGNITION-BASED ALGORITHMS

In this section, we present and describe eight recognition-based algorithms that we are surveyed from 1999 till 2009, especially on their lacks.

*D. Passface Scheme*

In 2000, this method developed which used faces as an object for password. During enrolment procedure, the users select whether their Passface consist of male or female picture. Then they choose four faces from the database as their future password. On the next step, a trial version starts for user in order to learn the real login process. During trial, the users taken twice through the Passface login procedure with their Passface which is shown to them. The enrolment will be completed by correctly identifying their four Passfaces twice in a row with no prompting, entering an enrolment password.

During login phase which is been shown in Fig 1, a grid that contain 9 pictures is shown to the user. This grid only contains one of the user's passwords, and the other eight pictures are selected from the database. As the users' password contain four faces, so the grid is shown four times. However, no grid contains faces found in the other grids, and the order of faces within each grid is randomized. These features help secure a user's Passface combination against detection through shoulder-surfing and packet-sniffing [20].

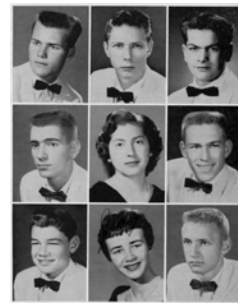

Figure III1: Passface Scheme, 1999

In 2004, a research showed that users selected attractive faces or faces of their own race more than others. The gender and attractiveness of the faces also bias password choice. In this research twelve categories of different faces gathered include, typical Asian males, typical Asian females, typical black males, typical black females, typical white males, typical white females, Asian male models, Asian female models, black male models, black female models, white male models and white female models. The results showed that Asian females and white females chose from within their race roughly 50% of the time; while males chose whites over 60% of the time, and black males chose blacks roughly 90% of the time [22, 25]. This makes the password easier to guess and suggested, so the Passface scheme can be vulnerable to guessing attack.

In 2006, a research compares the vulnerability of Passface with reference to its keyboard or mouse usage. The results showed that switching the configuration from mouse input to keyboard input decrease the vulnerability to shoulder surfing attack. As the attacker needs to look in two places at the same time, Passface with keyboard were more resistant to shoulder surfing attack [27].

Although most graphical password are resistance to being written down or verbally share, a research in 2008 focused on Passface in order to analysis its vulnerability to password description. Password description is any non digital attempt to record a password using verbal or non verbal means. In this research, during login process, an audio description of one of the face was played for the user. Then the designer used three different methods for selecting the eight decoy images. The three different methods are showed separately in Fig 2.

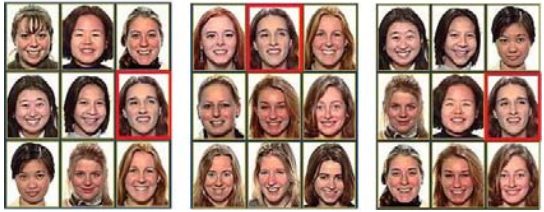

Figure 2: Three Different Methods in Selecting 8 Decoy Images
(Method 1, Method 2, And Method 3)

Method 1: The decoy images selected randomly by the system from a set of same age sex.

Method 2: The decoy images selected by visual similarity to the password face. To understand which image looked more similar, a group of contributors judge and decided on the resemblance of images.

Method 3: The decoy images selected based on the similarity to verbal description of the password face and the decoys.

The results showed that in random group eight people were successful among fifty-four attempts. The visual group showed four successful attempts login among fifty-six attempts and in the last condition only one person was successful among fourty-nine attempts. Therefore, it can be seen Passface can verbally describe but the correct choice of decoys can reduce the vulnerability of Passface to description. The decoy images should not have any significant characteristic either relating to the person (like race) or their face (like hair color or length, nose or ear size and shape) in order to make it hard for the user to describe it for the others [31]. According to [18] the Passface scheme can successfully map with several usability features of ISO standard like easy to use, easy to create and easy to recognize.

There are several drawbacks with Passface. Firstly, usage of keyboard or mouse could affect the threats of shoulder surfing attackers. The shoulder surfer has more chance if the Passface designer use mouse as an input device [27]. On the other hand, the research showed that performance like the time to complete the authentication process is slower than textual passwords, because users have to pass through a number of faces [6]. According to the images which are used by the designer, this scheme could be vulnerability to guessing attack [22, 25].

*E. Déjà vu Scheme*

This model proposed in 2000, by letting users to select specific number of pictures among large images portfolio. As the designer wanted to reduce the chance for description attack, the pictures create according to random art (one of the hash visualization algorithm). Firstly, one initial seed (a binary string) is given and then one random mathematical formula generates which defines the color value for each pixel in image. According to Fig 3, the output will be one random abstract image. Because the image depends only on the initial seed, it is not necessary to store the images pixel by pixel so only the seeds need to be stored in the trust server. During authentication phase, the user should pass thought a challenging set which his portfolio mixes with some decoy images. If the user can identify his entire portfolio successfully, he will be authenticating [21].

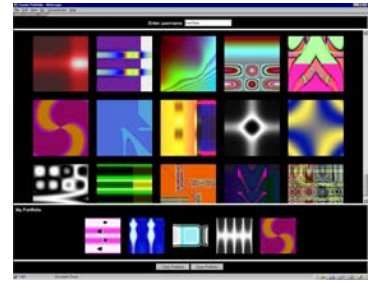

Figure.3: Déjà vu Scheme, 2000

There are several drawbacks with this method. Firstly the creation of portfolio in this method need about sixty second for the user which is longer than the time need for creating textual password (twenty five second). The login phase in this method takes longer time for the user in the compare to textual password [21]. On the other hand the process of selecting a picture from database can be tedious and consuming for the user. Another drawback could be the need for saving the seeds in the portfolio images of each user in plain text. [5].

*F. Triangle Scheme*

In 2002, Triangle algorithm proposed by a group which created several numbers of schemes which can overcome shoulder surfing attack. Their first scheme named, Triangle which is shown in Fig 4. In this method, the system randomly put a set of N objects which could be a hundred or a thousand on the screen. In addition, there is a subset of K objects previously chosen and memorized by the user. In other words, these K objects are the user passwords.

During login the system will randomly select a placement of the N objects then the user must find three of his password objects and click inside the invisible triangle created by those three objects or click inside the convex hull of the pass objects that are displayed. In addition, for each login this challenge is repeated a few times using a different display of some of the N objects. Therefore, the probability of randomly clicking in the correct region in each challenge is very low [23].

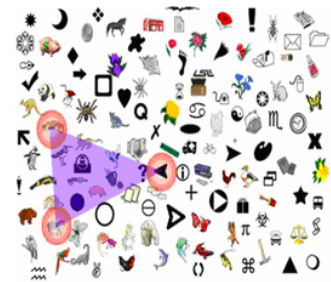

Figure 4: Triangle Scheme, 2002

*(IJCSIS) International Journal of Computer Science and Information Security,*
*Vol. 6, No. 2, 2009*

There are several drawbacks with this method. The designer of this method suggest usage of one thousand objects in the login phase so that the method could be resistant to shoulder surfing attack. But the problem is that the usage of this amount of object make the display very crowded and the objects almost indistinguishable. However, by using fewer objects, it will lead to a smaller password space and cause the resulting convex hull be large [5].

### G. Movable Frame Scheme

In 2002, this model produced using the same ideas and assumptions as Triangle scheme with the same designers. In this method the user must locate three out of K objects which these three are user passwords. As it is shown in Fig 5, only three pass objects are displayed at any given time and only one of them is placed in a movable frame.

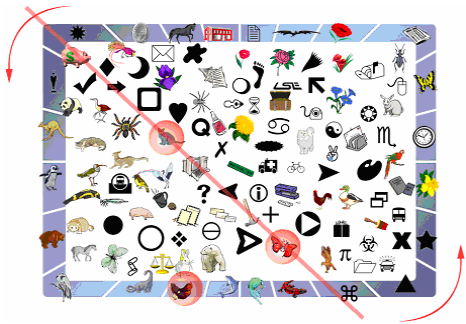

Figure 5: Moveable Frame Scheme, 2002

During login phase the user have to move the frame and the objects within it by dragging the mouse around the frame until the password object placed on the frame lines up with the other two pass objects. To minimize the likelihood of randomly moving the frame, the procedure is repeated a few times [23]. The drawback with this algorithm is that the process is unpleasant, confusing and time consuming since there are too many objects [27].

### H. Picture Password Scheme

In 2003, this algorithm designed especially for handheld device like Personal Digital Assistant (PDA). As it is shown in Fig 6, during enrollment, a user selects a theme identifying the thumbnail photos to be applied and then registers a sequence of thumbnail images that are used as a future password. When the PDA is turn on, the user must enter the current enrolled image sequence for verification to gain access to the device. After a successful authentication, the user may change the password and selecting a new sequence or theme.

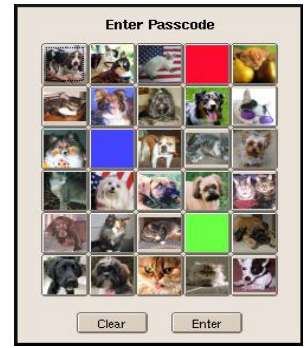

Figure 6: Picture Password Scheme, 2003

As the numbers of thumbnail photos are limited only to 30, the size of the password space is considered small. So to ensure that the password space is comparable to alphanumeric one, the designer added the second method of selecting thumbnail element. Besides selecting individual thumbnail elements as before, one could select two thumbnail elements together to compose a new alphabet element. This was done by using a shift key to select uppercase or special characters on a traditional keyboard, but in this context each thumbnail element serves as a shift key for every other element, including itself. With this addition, the password space expands from thirty elements to nine hundred and thirty elements, which compares favorably to the ninety five printable ASCII characters available from a traditional keyboard. However, this will make the memorability of the created password become more complex and difficult [26].

The drawback of this model is that the addition of shift key causes the algorithm complex and difficult.

### I. Man et al. Scheme

In 2003, this algorithm proposed as a new method for graphical password shoulder surfing resistant. In this algorithm all the pictures have assigned a unique code. As it is shown in Fig 7, during authentication the user is challenged with several scenes which contain several password objects and many decoy one. As there is a unique code for each password object, the user will enter the string of code for his password.

It is very hard for shoulder surfer to crack this kind of password even if the whole authentication process is recorded. However, this method still requires users to memorize the code for each password object variant. For example, if there are 4 pictures each with 4 variants, then each user has to memorize 16 codes.

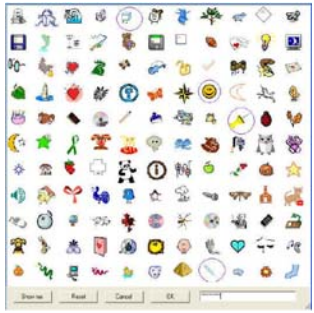

Figure 7: Man et al Scheme, 2004

The drawback of this method is that, although the password objects provide some cues for recalling the codes, it is inconvenient for the user to memorize all passwords with different cases.

*J. Story Scheme*

In 2004, the story scheme proposed by categorizing the available picture to nine categories which are animals, cars, women, food, children, men, objects, nature and sport. According to Fig 8, the users have to select their passwords from the mixed pictures of nine categories in order to make a story easily to remember. There were some users who used this method without defining a story for themselves [22].

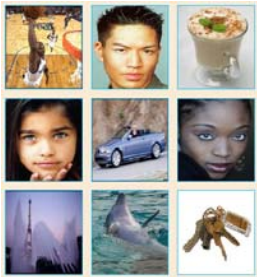

Figure 8: Story Scheme, 2004

This research showed that the story scheme was harder to remember in compare to Passface authentication.

*K. Jetafida Scheme*

In 2008, this model is proposed based on trying to gather all the usability features, like ease of use, ease to create, ease to memorize, ease to learn and acceptable design and layout in one algorithm. According to Fig 9, during registration, the user will select three pictures as a password and then sort them according to the way he wanted to see them in login phase.

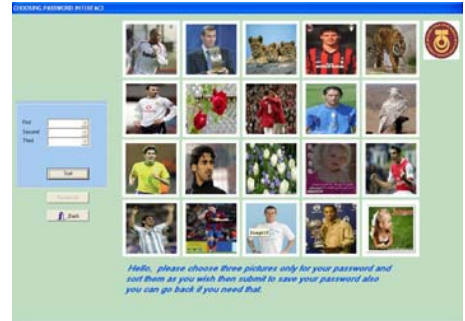

Figure 9: Jetafida Scheme, 2008

In login phase the password of the user will mixed with seven teen color pictures to create more usability. Around thirty people participate in trial version. According to them, 40% believed the algorithm is ease to use, 50% believed on ease to creation, 55% found the new algorithm easy to memorize, 57% user agree the algorithm is easy to learn and at last around 53% found the design and screen layout acceptable [24]. As the algorithm is very new, there is no special drawback in any survey until now.

IV. USABILITY IN ISO STANDARD

International Organization for Standardization (ISO) is the world's largest developer and publisher of International Standards. The ISO developed a variety of models to measure usability, but these models cannot cover totally in any of schemes. Among all different ISO methods, three of them describe the usability and its features in details. In the following section we discuss ISO 9241, ISO 9126 and ISO 13407 and then present a comparison table for the usability of graphical schemes.

*A. ISO 9241*

ISO 9241 is a series of international standards of ergonomics requirements for office work with visual display terminals. It provides requirements and recommendations concerning hardware, software and environment attributes that contribute to usability. According to ISO 9241-11, the usability definition is [17]:

*"Extent to which a product can be used by specified users to achieve specified goals with effectiveness, efficiency and satisfaction in a specified context of use."*

There are seventeen different parts in this ISO under four categories. In "general category" two different parts are mentioned. Under "material requirement category" and "environment category", parts three to nine exist, which deal with hardware design requirements and guidelines that can have implications on software. Parts ten to seventeen are in the "software category" which deal with software attributes [17].

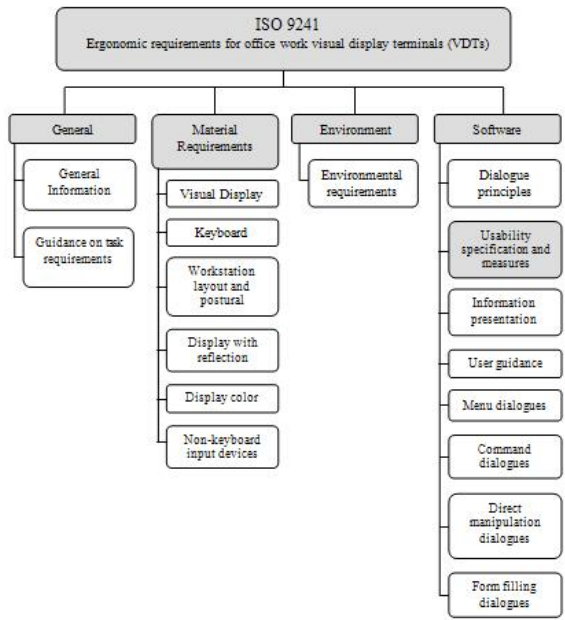

Figure 10: The ISO 9241

As we can see from Fig 10, among different part of ISO 9241, the part eleven defines the usability from three main components which are:

*Effectiveness:* Describes the interaction from a process point of view or accuracy and completeness with which users achieve specified goals. Means how well do the users achieve the goals they set out in using the system?

*Efficiency:* Resources expended in relation to the accuracy and completeness with which users achieve goals.

*Satisfaction:* Freedom from discomfort, and positive attitudes to the use of the product. Refers to a user point of view or how the users feel about their use of the system?

This part explains how to identify the information that it is necessary to take into account when specifying usability in terms of measures of user performance and satisfaction. ISO 9241-11 recommends a process oriented approach for usability, by which the usable interactive system is achieved through a human centered design process.

B. ISO 9126

ISO 9126 address software quality from the product point of view. In fact this is the most extensive software quality model which presents quality as a whole set of characteristics. According to Fig 11, this ISO divides software quality into six general categories which are: functionalities, reliability, usability, effectiveness, maintainability and portability [35]. Part three of ISO 9126 defined the usability as:

"A set of attributes that bear on the effort needed for use and on the individual assessment of such use, by a stated or implied set of users".

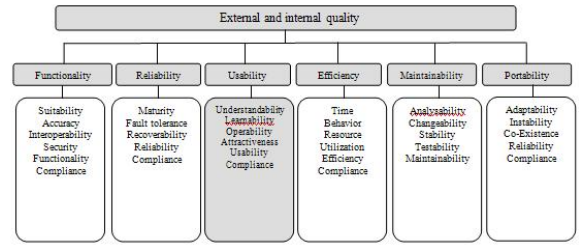

Figure 11: The ISO 9126

Usability was seen like an independent factor of software quality. It treated software attributes, mainly its interface that makes it easy to use. As shown in Fig 11, the major attributes consist of understandability, learnability, operability and attractiveness [35].

C. ISO 13407

This is an ISO standard focus on human centered design in order to create interactive system development for making the system more usable. The application of human factor enhances effectiveness, efficiency and human working condition. In order to cover this aim, this ISO emphasis on learnability of user which provides more productivity and quality to the system. The definition of usability, effectiveness, efficiency and satisfaction in this standard is referred to ISO 9241-11 definition and are the same. Here we mention some of the benefits of adding more learnability features to human design [32].

- The system will be easier to understand and use, thus reducing training and support costs.
- Improve user satisfaction and reduce discomfort and stress.
- Improve the productivity of users and the operational efficiency of organizations.

Based on the survey done regarding the recognition-based algorithms and the three ISO standards which described in the previous section, we develop Table I as shown below which concatenate all these usability attributes. For creating this table we focus on ISO usability component which are effectiveness, efficiency and satisfaction. Then the attributes of each of them create based on previous research, for instance using mouse cause the user to be more satisfy in compare with keyboard usage.

TABLE I: USABILITY ATTRIBUTES FROM ISO STANDARDS

| Usability features | Attributes | Attributes for GUA | Abbreviation |
|---|---|---|---|
| Effectiveness | Reliability & Accuracy | Reliability & Accuracy | R&A |
| Efficiency | The Utilization in Real World | Applicable | Applicable |
| Satisfaction | Easy to Use | Use the Mouse Easily | Mouse Usage |
| | Easy to Create | Select Simple Way to Create the Password | Create Simply |
| | Easy to Memorize | Meaningful | Meaningful |
| | | Memorability | Memorability |
| | Easy to Execute | Select Simple Steps of Registration and Login | Simple Steps |
| | Good View | Select Good Interface | Nice Interface |
| | Easy to Understand | Simple Training Session | Training Simply |
| | Pleasant | Pleasant Picture | Pleasant Picture |

## V. USABILITY COMPARISON ON RECOGNITION-BASED ALGORITHMS

There are several surveys which concentrate on usability of different graphical scheme which the latest one related to one survey which has been done by University Technology Malaysia [18]. According to this research, the items of usability are easy to use, easy to create, easy to memorize and easy to learn. As the techniques which is been used by recognition, pure and cued schemes are different, so the features in their usability table is not the same. For example, in recognition-based scheme the usage of different pictures is one of the main items. Therefore, the "pleasant picture" is a meaningful item in their usability table. The creation of three following tables are based on [18, 19] surveys.

TABLE II: THE USABILITY FEATURES IN RECALL-BASED TECHNIQUES

| Row | Recognition based schemes | Usability Features | | | | | | | | Efficiency | Effectiveness |
|---|---|---|---|---|---|---|---|---|---|---|---|
| | | Satisfaction | | | | | | | | | |
| | | Mouse usage | Create Simply | Meaningful | Assignable Image | Memorability | Simple Steps | Nice Interface | Training Simply | Pleasant Picture | Applicable | R&A |
| 1 | PassFace | Y | Y | N | Y | Y | Y | Y | Y | N | Y |
| 2 | Déjà vu | Y | Y | N | Y | N | Y | N | Y | N | Y |
| 3 | Triangle | Y | Y | N | N | Y | Y | N | N | N | Y |
| 4 | Movable Frame | Y | Y | N | N | Y | Y | N | N | N | Y |
| 5 | Picture Password | Y | Y | Y | Y | Y | Y | Y | Y | Y | N |
| 6 | Story | Y | Y | Y | Y | Y | Y | N | Y | Y | N |
| 7 | Man | N | Y | N | Y | Y | Y | N | Y | Y | N |
| 8 | Jetafida | Y | Y | N | Y | Y | Y | Y | Y | Y | N |

Notes: Y: Yes; N: No

## VI. COMMON ATTACKS IN GRAPHICAL PASSWORD SCHEMES

The graphical password schemes cover common attacks are defined in the following section.

### Password Brute Forcing Attack

In this attack which has the attack pattern ID 112, the attacker tries every possible value for a password until they succeed [34]. A brute force attack, if feasible computationally, will always be successful because it will essentially go through all possible passwords given the alphabet used and the maximum length of the password. A system will be vulnerable to this type of an attack if it does not have a proper mechanism to ensure that passwords are strong passwords that comply with an adequate password policy. In practice, a pure brute force attack on passwords is rarely used, unless the password is suspected to be weak. The speed which an attacker discovers a secret is directly related to the resources that the attacker has. This attack method is resource expensive as the attackers' chance for finding user's password is high only if the resources be as complete as possible.

### Dictionary Based Password Attack

In this attack which has the attack pattern ID 16, an attacker tries each of the words in a dictionary as passwords to gain access to the system via some user's account [34]. If the password chosen by the user was a word within the dictionary, this attack will be successful. This is a specific instance of the password brute forcing attack pattern.

### Guessing Attack

As many users try to select their passwords based on their personal information like the name of their pets, passport number, family name and so on, the attacker try to guess password by trying these possible password. Password guessing attacks can be broadly categorized into online password guessing attacks and offline dictionary attacks. In an online password guessing attack, an attacker tries a guessed password by manipulating the inputs of one or more oracles. In an offline dictionary attack, an attacker exhaustively searches for the password by manipulating the inputs of one or more oracles [30].

### Spyware Attack

Spyware is a type of malware which installed on computers with the aim of collecting sensitive information of users, using a key logger or key listener. This information gathered without user's knowledge and report back to an outside source. During graphical password authentication the attacker attempt to gain sensitive information like user names or selected passwords images by intercepting information exchanged.

### Shoulder Surfing Attack (or Observer Attack)

Shoulder surfing refers to using direct observation techniques, such as looking over someone's shoulder, to get information. Shoulder surfing is effective in crowded places because it is really easy to stand near someone and watch them entering a PIN number at an ATM machine. This attack is also possible at a distance using vision-enhancing devices like miniature closed circuit cameras which can be concealed in ceilings, walls or fixtures to observe data entry. To prevent shoulder surfing, it is advised to shield paperwork or the keypad from view by using one's body or cupping one's hand. Nearly most graphical password schemes are quite vulnerable to shoulder surfing.

*Social Engineering Attack*

In this kind of attack, an attacker uses human interaction to obtain or compromise information about an organization or computer systems, so he claimed to be one of employee in order to gain identity. On the other hand, the attackers try to ask many questions in order to infiltrate an organization's security. If an attacker is not able to gather enough information from one source, he or she may contact another source within the same organization and rely on the information from the first source to add to his or her credibility.

In the following section, we create the comparison table (Table III) for these attacks based on the surveys [5, 11 and 30]. As we can see in this table three methods named, picture password, story and Jetafida do not have any study on their reactions toward common graphical password attack. For Jetafida method it may be because this method is new and the evaluation on the attacks have not done yet. For two other methods the previous survey did not mention any thing about their weakness toward attack. So these items which are not filled will be our future work.

TABLE III: THE ATTACK COMPARISON IN PURE AND CUED BASED SCHEMES

| Row | Recognition Based Schemes | Attacks | | | | | |
|---|---|---|---|---|---|---|---|
| | | Brute Force | Dictionary | Guessing | Spyware | Shoulder surfing | Social Engineering |
| 1 | PassFace | Y | Y | Y | N | Y | N |
| 2 | Déjà vu | Y | N | Y | N | Y | N |
| 3 | Triangle | Y | N | Y | N | N | N |
| 4 | Movable Frame | Y | N | Y | N | N | N |
| 5 | Picture Password | - | - | - | - | Y | - |
| 6 | Story | - | - | - | - | Y | - |
| 7 | Man | Y | N | N | Y | N | N |
| 8 | Jetafida | - | - | - | - | Y | - |

Notes: Y: Yes; N: No

VII. CONCLUSIONS

In this study, eight algorithms from recognition-based graphical password authentication are reviewed and surveyed. During our research, we identify several drawbacks which can cause attacks. Therefore, it can be concluded that the common drawbacks on these eight algorithms were:

- Users were fascinated by the pictures which drawn by other users, so frequently we can see the common picture for password.
- The users can hardly remember the sequence of drawing after period of time.
- The users tend to select the weak passwords which are vulnerable to the graphical dictionary attack.
- Not all the users are familiar with using mouse as a drawing input device for graphical password.
- The memorize ability and usability of some of the algorithms are difficult.
- The users tend to select the weak passwords which can cause the password to be guessable or predictable.

After explained completely eight Recognition-Based graphical authentication algorithms by mentioning their description, lacks and attacks, we survey on ISO standards for usability features and collect the major usability features of usability from three ISO standards. Then, try to survey on attack patterns and define common attacks for graphical user authentication methods. Finally we make a comparison table among Recognition-Based algorithms based on ISO usability attributes and Attack Patterns.

ACKNOWLEDGMENT

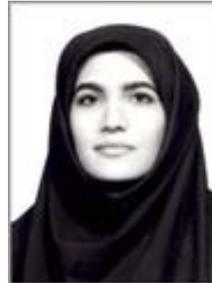
**Farnaz Towhidi** obtained her Bachelor degree in Software Engineering from Azad University of Tehran, Iran. She followed by Master degree of Computer Science - Information Security in Centre for Advanced Software Engineering, University technology of Malaysia. She has four papers in various international and national conferences.

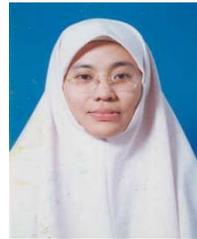
**Maslin Masrom** received the Bachelor of Science in Computer Science (1989), Master of Science in Operations Research (1992), and PhD in Operations Mgmt/Mgmt Info System (2003). She is an Associate Professor, Department of Science (Computer Science Unit), College of Science & Technology, University Technology Malaysia *International Campus,* Kuala Lumpur. Her current interests include information security, ethics in computing, e-learning, structural equation modelling and information technology/information system management. She has published articles in both local and international journals such as Information and Management Journal, Oxford Journal, Journal of US-China Public Administration, MASAUM Journal of Computing, ACM SIGCAS Computers & Society and International Journal of Cyber Society and Education.